\documentclass[fleqn,usenatbib]{mnras}

\usepackage{mathptmx}
\usepackage{txfonts}

\hypersetup{pdfauthor={S. F. Hoenig},
		   pdftitle={Cosmology with AGN dust time lags -- Simulating the new VEILS survey},
		   pdfkeywords={galaxies: active -- distance scale -- infrared: galaxies},
		   bookmarksnumbered=true}

\volume{in press}
\usepackage[T1]{fontenc}
\usepackage{ae,aecompl}

\usepackage{graphicx}	% Including figure files
\usepackage{amssymb}	% Extra maths symbols

\newcommand{\veils}{\textit{VEILS}}

\title[VEILS AGN Dust Cosmology]{Cosmology with AGN dust time lags -- Simulating the new VEILS survey}

\author[S. F. H{\"o}nig et al.]{S. F. H{\"o}nig,$^{1}$\thanks{Contact e-mail: \href{mailto:S.Hoenig@soton.ac.uk}{S.Hoenig@soton.ac.uk}}
D. Watson,$^2$ M. Kishimoto,$^3$ P. Gandhi,$^1$ M. Goad,$^4$ K. Horne,$^5$ 
\newauthor F. Shankar,$^1$ M. Banerji,$^{6,7}$ B. Boulderstone,$^1$ M. Jarvis,$^{8,9}$ M. Smith,$^1$ M. Sullivan$^1$
\\
$^1$ Department of Physics \& Astronomy, University of Southampton, Highfield Campus, Southampton SO17 1BJ, UK \\
$^2$ Dark Cosmology Centre, Niels Bohr Institute, University of Copenhagen, Juliane Maries Vej 30, DK-2100 Copenhagen \O, Denmark \\
$^3$ Department of Physics, Faculty of Science, Kyoto Sangyo University, Kamigamo-motoyama, Kita-ku, Kyoto 603-8555, Japan \\
$^4$ Department of Physics and Astronomy, College of Science and Engineering, University of Leicester, University Road, \\
\hspace{0.15cm} Leicester LE1 7RH, UK \\
$^5$ SUPA Physics and Astronomy, University of St Andrews, KY16 9SS Scotland, UK \\
$^6$ Institute of Astronomy, University of Cambridge, Madingley Road, Cambridge CB3 0HA, UK \\
$^7$ Kavli Institute for Cosmology, University of Cambridge, Madingley Road, Cambridge CB3 0HA, UK \\
$^8$ Oxford Astrophysics, Department of Physics, Keble Road, Oxford OX1 3RH, UK \\
$^9$ Physics Department, University of the Western Cape, Cape Town 7535, Republic of South Africa}

\date{Accepted 2016 September 27. Received 2016 September 15; in original form 2016 July 14}

\pubyear{2016}

\begin{document}
\label{firstpage}
\pagerange{\pageref{firstpage}--\pageref{lastpage}}
\maketitle

\begin{abstract}
The time lag between optical and near-infrared continuum emission in active galactic nuclei (AGN) shows a tight correlation with luminosity and has been proposed as a standardisable candle for cosmology. In this paper, we explore the use of these AGN hot-dust time lags for cosmological model fitting under the constraints of the new VISTA Extragalactic Infrared Legacy Survey \veils. This new survey will target a 9 deg$^2$ field observed in $J$- and $Ks$-band with a 14-day cadence and will run for three years. The same area will be covered simultaneously in the optical $griz$ bands by the Dark Energy Survey, providing complementary time-domain optical data. We perform realistic simulations of the survey setup, showing that we expect to recover dust time lags for about 450 objects out of a total of 1350 optical type 1 AGN, spanning a redshift range of $0.1 < z < 1.2$. We use the lags recovered from our simulations to calculate precise distance moduli, establish a Hubble diagram, and fit cosmological models. Assuming realistic scatter in the distribution of the dust around the AGN as well as in the normalisation of the lag-luminosity relation, we are able to constrain $\Omega_\Lambda$ in $\Lambda$CDM with similar accuracy as current supernova samples. We discuss the benefits of combining AGN and supernovae for cosmology and connect the present work to future attempts to reach out to redshifts of $z>4$. 
\end{abstract}


%%%%%%%%%%%%%%%%%%%%%%%%%%%%%%%%%%%%%%%%%%%%%%%%%%

%%%%%%%%%%%%%%%%% BODY OF PAPER %%%%%%%%%%%%%%%%%%

\section{Introduction}

Arguably 10 per cent of all large galaxies host an active galactic nucleus (AGN) in their centre. Given their high luminosities, AGN can be detected from low-redshift out to the early universe at $z>7$. Moreover, on human time scales, AGN are rather predictable with only a few sources turning on or off completely \citep[e.g.][]{Kee12a,Kee12b}. These traits would make them desirable tools for cosmology if their emission were standardisable.

Several routes are currently being pursued to establish AGN as standard candles. \citet{Wat11} showed that the known relation between the size of the broad emission line region and AGN luminosity can be used to constrain cosmological parameters \citep[see also][]{Haa11,Cze13,Kin14}. The sizes of these structures are determined using the time lags between the incident radiation and the reaction in the reprocessed emission from the observed region. Alternatively to emission lines, the lag between the optical continuum and the dust continuum correlates with luminosity and can also serve as a standard candle \citep[e.g.][]{Okn99,Okn01,Hon14a,Yos14}. Beyond this, it was proposed to use either the emission line time lags \citep[in combination with interferometry;][]{Elv02}, lag profiles \citep[based on photoionisation modelling;][]{Hor03} or the dust lags \citep[with infrared interferometry;][demonstrated for NGC 4151]{Hon14b} as standard rulers, which set an absolute distance scale to AGN and directly measure the Hubble constant $H_0$.

The major advantages of dust time lags as compared to emission line lags are (1) their tighter relation between lag and luminosity, when considering the same set of objects \citep[e.g.][]{Kos14}, and (2) the use of photometry instead of spectroscopy. On the other hand, broad emission line lags can be measured out to redshift 4 or even beyond \citep[e.g.][]{Wat11,Cze13,Kin14,Kin15}. When combined, both AGN standardised candles will explore a large region in redshift space.

In this paper, we provide the scientific motivation, foundation, and simulations for using AGN hot-dust time lags as a standard candle in the redshift range $0.1 < z < 1.2$. This range is motivated by the fact that ground-based facilities with $\sim2\micron$ imaging capability can be used for a systematic survey. We focus on the design of the VISTA Extragalactic Infrared Legacy Survey \veils, a new ESO public survey scheduled to run on the 4m VISTA telescope for 3 years starting 2017. In the next section, we will briefly review the physical background of the size-luminosity relation in the near-infrared (IR) and discuss some characteristics as constrained by observations. In Sect.~\ref{sec:distmod}, we will derive the equations to turn lags into distance moduli, which are required to establish a Hubble diagram and fit cosmological models. In Sect.~\ref{sec:sim} we present our simulations of \veils\ AGN light curves and discuss the methods employed to recover time lags. In Sect.~\ref{sec:res} we show the results of our simulations and quantify the constraints on cosmological parameters we can expect from the survey, together with a comparison to type Ia supernovae. Practical challenges and strategies are outlined in Sect.~\ref{sec:chal}. Finally, we summarise our findings and present the broader context and legacy of the \veils\ AGN variability survey in Sect.~\ref{sec:sum}.

\section{The lag-luminosity relation in the near-IR}\label{sec:lag_lum}

The multi-wavelength emission of AGN shows a prominent IR bump starting at about 1\,$\micron$ and extending to 100-1000\,$\micron$ with a peak in the 20-30\,$\micron$ range\footnote{Note that in many AGN-hosting galaxies, the IR emission peaks at much longer wavelengths. These peaks are associated with star formation in the host galaxy or nuclear region rather than the AGN itself. While the star formation contamination might be an issue in the mid- and far-IR, it does not affect the near-IR emission, in particular the variable component seen from the AGN.}. This emission is associated with thermal reprocessing of the ultraviolet (UV) and optical radiation from the ``big blue bump'' (BBB; or accretion disk) much closer to the black hole (dust-to-BBB size ratio $\ga$20). 

Dust can only survive temperatures up to about $T_\mathrm{sub}\sim$1500\,K, above which the grains sublimate. This implies that the near-IR emission at $1-2\,\micron$ delineates a sharp boundary between dust-free and dust-containing gas, with its exact size set by the (solid-state) properties of the dust grains \citep[chemical composition, size; e.g.][]{Bar87,Phi89}. Thus, using the radiative transfer equations in thermal equilibrium, we can relate the AGN luminosity $L_\mathrm{AGN}$ to the hot dust emission close to the sublimation radius $r_\mathrm{sub}$ as \citep{Hon11}
\begin{equation}
L_\mathrm{AGN} = 16 \pi r_\mathrm{sub}^2 f_\mathrm{abs}^{-1} \ Q_\mathrm{abs;P}(T_\mathrm{sub}) \ \sigma_\mathrm{SB} T_\mathrm{sub}^4,
\end{equation}
where $f_\mathrm{abs}$ is the fraction of incident AGN flux absorbed per dust particle, $\sigma_\mathrm{SB}$ is the Stefan-Boltzmann constant, and $Q_\mathrm{abs;P}(T_\mathrm{sub})$ is the \textit{normalised} Planck mean absorption efficiency of the dust for given sublimation temperature $T_\mathrm{sub}$, i.e. $Q_\mathrm{abs;P}(T_\mathrm{sub}) = \int Q_\mathrm{abs;\nu} \pi B_\nu(T_\mathrm{sub}) \mathrm{d}\nu / (\sigma_\mathrm{SB} T_\mathrm{sub}^4)$. Most importantly, $f_\mathrm{abs}$ and $Q_\mathrm{abs;P}$ are related with both parameters approaching unity for large dust grains, which emit very similarly to a black body. 

We can now simplify this equation by absorbing all constant parameters (including $T_\mathrm{sub}$, $Q_\mathrm{abs;P}$ and $f_\mathrm{abs}$) into a normalisation parameter $k_n$. After replacing $r_\mathrm{sub}$ with the corresponding time lag $\tau_\mathrm{sub} = r_\mathrm{sub}/c$, we obtain
\begin{equation}\label{eq:ltau}
L_\mathrm{AGN} = k_n \cdot \tau_\mathrm{sub}^2.
\end{equation}
In this notation, the actual object-to-object differences in hot dust composition, geometry, and global distribution are reflected by the observed scatter $\sigma_k$ in the normalisation of the lag-luminosity relation of an AGN sample. Interestingly, this scatter is not very large. When using the $V$-band as a proxy for $L_\mathrm{AGN}$ and observing lags between the $V$- and $K$-bands, \citet{Kos14} find $\log \sigma_k = 0.14-0.16$ dex (see their Table 9), which implies similar overall characteristics of the hot-dust emitting region in their sample of 17 AGN (see also Appendix~\ref{sec:app_taulum}). These similarities can be easily understood: First, when observing at wavelengths equivalent to temperatures of 1500\,K or higher, the emission is restricted to grains that can actually survive such high temperatures. Indeed, silicate dust grains already sublimate at lower temperatures of $800-1200$\,K while carbonaceous grains can get hotter \citep[e.g.][]{Phi89}. Second, when observing fluxes from a relatively confined region, the emission will be biased towards grains with higher emissivity (= closer to black body radiation), which are typically the larger grains within the composition. Therefore, the emission at wavelengths close to the (carbonaceous) sublimation temperature will be dominated by black-body-like dust grains (large, carbon) regardless of the details of the original ISM composition. This is strongly supported by surface emissivity measurements of nearby AGN in the $K$-band using IR interferometry \citep{Kis11b}.

\section{Using dust time lags as standard candles}\label{sec:distmod}

Both the observational evidence and the rather simple radiation and solid-state physics at work underline the promise of using the hot dust lag-luminosity relation for cosmological applications. For this, we have to turn the theoretical framework outlined in the previous section into an observational tool. The common function to test the standard model in cosmology is the relation between distance and redshift. Given the lag-luminosity relation, we need to determine the luminosity distance $D_L = (1+z) \cdot c/H_0 \ \int \mathrm{d}z/E(z)$ of an AGN and compare it to its measured redshift $z$. Here, $H_0$ is the Hubble constant and $E(z) = (\Omega_m(1+z)^3 + \Omega_\Lambda(1+z)^{3(1+w)})^{1/2}$ (assuming a $w\Lambda$CDM cosmological model without curvature). $D_L$ is related to the distance modulus $m-M = 5 \log D_L/10\,\mathrm{pc}$ with the absolute magnitude $M$ to be determined from the lag-luminosity relation.

The lag-luminosity relation as written in eq.~(\ref{eq:ltau}) determines the bolometric AGN luminosity from the lag. In practice, it is difficult to measure the full SED of the AGN that contributes to dust heating. However, observations have shown that replacing the total luminosity with a monochromatic proxy in the restframe optical waveband regime provides small-enough scatter \citep[e.g][see also previous section]{Sug06,Kis07,Kos14}. We will follow the established convention and use the rest-frame $V$-band magnitude at 550\,nm as the proxy for $L_\mathrm{AGN}$ and measure the $Ks$-band to determine the hot dust time lag. Therefore, the standard candle relation for the hot-dust radius of AGN can be written as
\begin{equation}\label{eq:sc}
m_V - M_V^* = m_V + 2.5\log k_n + 5 \log \frac{\tau_K}{1+z},
\end{equation}
where $m_V$ is the rest-frame 550\,nm apparent magnitude and the factor $(1+z)^{-1}$ accounts for the relativistic time dilation of the lags. In this prescription, $k_n$ implicitly absorbs both the normalisation of the lag-luminosity relation as well as $H_0$ since they are essentially indistinguishable for practical purposes. $k_n$ has to be determined either from a low-redshift sample that does not distinguish between different cosmological models or by a global fit to a large sample. Since $k_n$ absorbs the absolute normalisation of the size luminosity relation, $M_V^*$ should not be considered as the true absolute magnitude but rather a close proxy for $M_V$, which is a common aspect of standardisable candles (e.g. supernovae).

We mention that there have been attempts to estimate $H_0$ from the normalisation of the AGN Hubble diagram using dust time lags \citep{Yos14}. In practise, this is rather difficult since it requires interpreting the normalisation of the lag-luminosity relation as a result of the dust grain emissivity at the sublimation radius. This involves using redshift-based luminosity distances to convert AGN flux to luminosity, thereby leading to a degeneracy between $H_0$ and emissivity. On the other hand, combining the dust time lags with near-IR interferometry allows measurement of direct distances to AGN \citep{Hon14b}, which can be used to independently measure $H_0$ and self-consistently calibrate the AGN hot-dust standard candle.

\section{Simulating the \veils\ transient survey for AGN}\label{sec:sim}

In this section, we will outline \veils\ and quantify the expected constraints on cosmological models by using AGN dust lags as obtained in the course of the survey. As demonstrated, \veils\ will allow us to observationally establish AGN dust lags as a new standardisable candle. With this new tool in hand, it will be possible to address tensions of low- and high-redshift constraints on cosmological parameters independently \citep[e.g.][and references therein]{Rie16}. The survey simulations and strategies described in the following have been tested against observations and approximately reproduce the observed scatter of the lag-luminosity relation (see Appendix~\ref{sec:app_taulum}).

\subsection{Overview of \veils}\label{sec:surv}

The VISTA Infrared Extragalactic Legacy Survey \veils\ is a new public $J$ and $Ks$ band survey on the 4m VISTA telescope at ESO's Paranal Observatory (Chile). Rather than providing limits on total survey depths, the survey strategy emphasises symbiosis between cadenced observations with per-epoch depths required by transient science cases (AGN and supernovae) and combined survey depths suitable for a range of galaxy evolution science cases.

\veils\ is scheduled to run over 3 years with approximately 6-month long observing seasons each year. We target 3\,deg$^2$ regions in each of the Chandra Deep Field South, Elias South field, and XMM deep field for a total of 9\,deg$^2$. The selected regions are simultaneously covered by the Dark Energy Survey (DES), which provides us with contemporaneous $griz$ light curves, complementing the $J$ and $Ks$ light curves from \veils. DES will observe each field about every 7 days while \veils\ will repeatedly visit each field every $10-14$ days. This will result in approximately 15 epochs per field per observing season or 45 epochs over 3 years in total. As demonstrated below, these kind of multi-band time-domain observations will allow for reconstructing enough AGN dust time lags to put competitive constraints on cosmological parameters. Table~\ref{tab:veils} summarises cadence and per-epoch depths for each filter. These parameters will be used for the survey simulations below. It is possible that DES will finish operations before the end of \veils\. In this case, we will complement the near-IR survey with optical observations to the same depth and similar cadence at the VLT Survey Telescope (VST), which is operated by ESO on the same site as VISTA.

\begin{table}
\caption{Survey parameters used as input to the simulations.}\label{tab:veils}
\begin{center}
\begin{tabular}{l c c c}
\hline
Survey & cadence & filter & 5$\sigma$ per-epoch depth \\
            &  (days)      &         &        (AB mag) \\ \hline
\veils & 14                & $Ks$ &  22.5 \\
     &                         & $J$ & 23.5 \\ \hline
DES & 7                   & $z$ & 24.0 \\
&                              & $i$ & 24.3 \\
&                              & $r$ & 24.1 \\
&                              & $g$ & 24.5 \\ \hline
\end{tabular}
\end{center}
\textit{Note:} The DES limits correspond to an analysis of the combined Year 1-3 deep transient fields, which are also targeted by \veils\ (M. Smith, priv. comm.).
\end{table}

\subsection{Defining an AGN mock sample}

\veils\ will cover a total area of 9 deg$^2$ for which we have to estimate the number of type 1 (=unobscured) AGN that can be detected. In order to create a mock sample of AGN with realistic parameters, we took the redshift-dependent optical luminosity function of \citet{Pal13} and estimated the number, luminosity distribution and redshift distribution of all unobscured AGN within the survey area. After accounting for the 5$\sigma$ detection limits of DES in $griz$ (see Table~\ref{tab:veils}), we are left with a sample of about 1350 type 1 AGN and we define these as our AGN mock sample. Note that since our sources are variable, the 5$\sigma$ limits refer to the long-term mean magnitude.

\subsection{Variability and survey simulations}\label{sec:varsim}

We follow previous work in \citet{Hon14a} to characterise the variability in the AGN mock sample and simulate optical and infrared light curves. This involves modelling the AGN variability as a stochastic correlated autoregressive (CAR) process \citep[e.g.][]{Kel09,Kel13}. For all the AGN in our sample, we assigned black hole masses to the luminosities drawn from the luminosity functions by assuming an Eddington ratio log-normal distribution centred at $\log l_\mathrm{Edd} = -1.2$ and a standard deviation of 0.3 dex (this covers Seyfert galaxies and quasars in the detectable luminosity range; see note at the end of this section). \citet{Kel13} show how the parameters of the CAR process correlate with observed AGN black hole masses and luminosities. We use these empirical relations, including their large scatters, to assign variability characteristics to the AGN mock sample. Based on these parameters, we simulate light curves for all of the bands and account for the observed dependence of variability amplitude on wavelength \citep[``bluer when brighter'' with amplitude $\propto \lambda^{-0.28}$;][]{Meu11}.

The simulated AGN ``big blue bump'' (BBB) light curves are used as input to a dust radiative transfer model to simulate the response of the dust emission to the AGN variability. We parameterise the dust distribution in terms of an observationally motivated projected radial power law $\propto r^a$ \citep[e.g.][]{Kis09a,Hon10,Kis11b} and an effective response amplitude $w_\mathrm{eff}$, which sets the fraction of dust/infrared light that actually varies \citep{Hon11}. We assign random parameters from a range of $0.25 < w_\mathrm{eff} < 0.85$ and $-2.5 < a < -0.5$ \citep[a detailed description of the process is available in the method section of][and references therein]{Hon14b}. In short, these parameters provide a realistic picture of how strongly the near-IR emission will react to the incident AGN variability and to what extent the variability will be smeared out.

\begin{figure*}
	\includegraphics[width=0.95\textwidth]{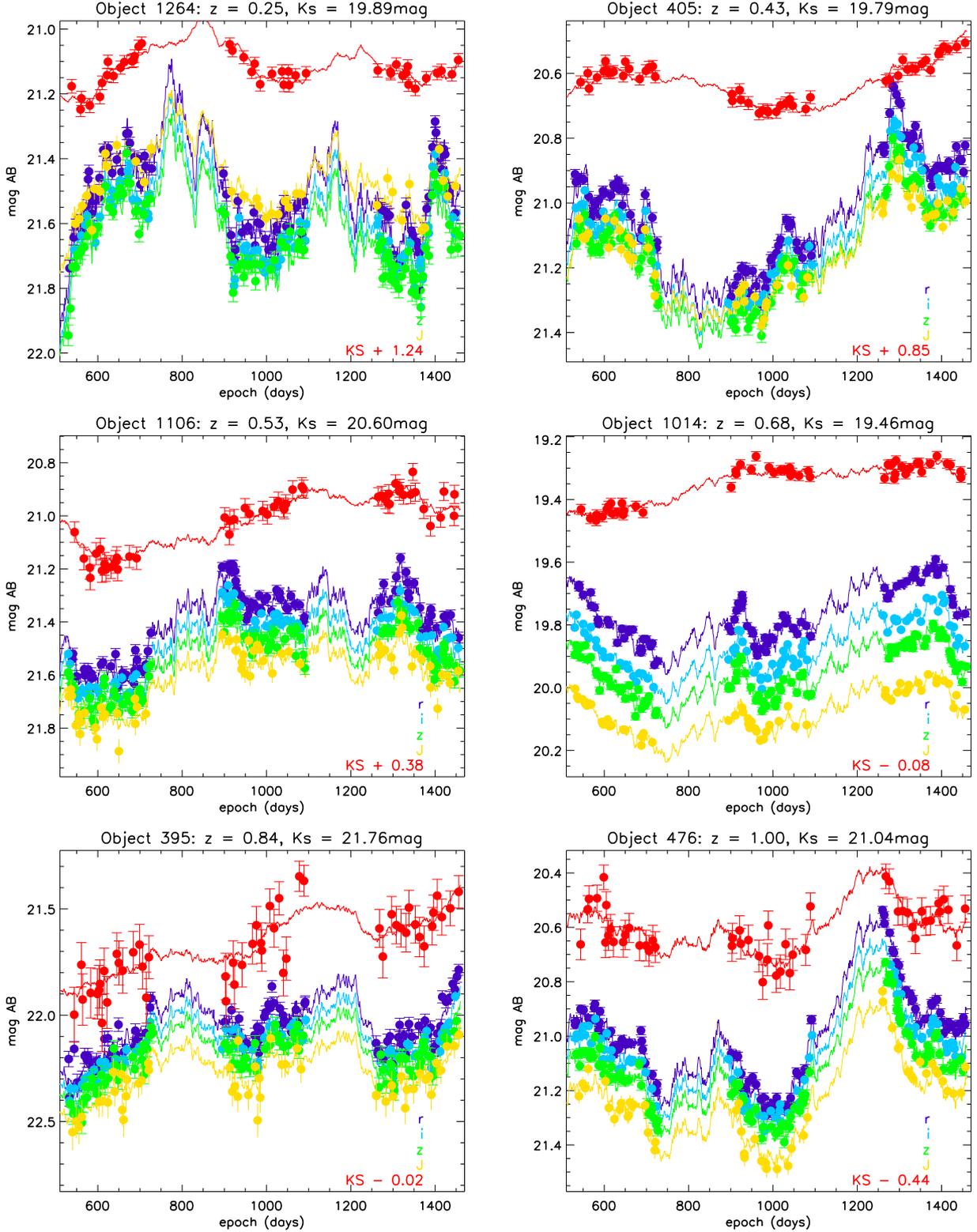}
	\caption{Simulated and mock-observed light curves for six objects in our mock AGN sample. The purple, blue and green dotted lines represent the input AGN light curves in the DES $riz$ bands while the yellow and red dotted lines are $J$- and $Ks$-band light curves in the VIRCAM bands ($Ks$ is shown as the upper-most curve in each panel). The filled circles with error bars represent the 3-year mock survey observations, which account for the limiting magnitudes in each band, the survey cadence, and the seasonal gaps. The $Ks$ band light curves are offset for illustration purposes and the offset is noted in each panel where applicable. The running numbers refer to the full type 1 AGN sample of 1350 objects.}
	\label{fig:ex_lc}
\end{figure*}

After combining the AGN and dust light curves in each filter, we apply the \veils\ and DES survey constraints listed in Table~\ref{tab:veils} to the light curves (see Sect.~\ref{sec:surv}), and account for the noise characteristics in the DES and \veils\ bands as well as some inhomogeneity in the observing cadence due to scheduling or bad observing conditions. In Fig.~\ref{fig:ex_lc} we show simulated light curves and mock observations for six AGN from the mock sample. The coloured lines represent the input model light curves for different bands, while the data points reflect the 3-year survey, including 6-month seasonal gaps, with a median distance between each epoch corresponding to the DES and \veils\ cadence. As illustrated by these examples, the significance of detecting a lag between optical and IR emission depends on various factors like variability characteristics, response/transfer function, and redshift (= restframe wavelength of the $Ks$-band).

We note that the exact shape of the $\log l_\mathrm{Edd}$ distribution does not affect the properties of the final AGN sample noticeably: In testing a log-uniform distribution with $-2 < \log l_\mathrm{Edd} < -0.4$, we obtain the same redshift and luminosity distribution of those objects that are recovered within the DES and \veils\ survey constraints.

\subsection{Recovering dust lags from the DES+\veils\ light curves}\label{sec:lag_recov}

With the mock survey observations in hand, we aim at recovering the dust lags as precisely as possible. The success rate mainly depends on the S/N of the observations, the cadence, and the fraction of dust emission varying in the $Ks$-band, which is a mixture of an AGN-intrinsic property (see Sect.~\ref{sec:varsim}) and redshift.

\begin{figure*}
	\includegraphics[width=0.95\textwidth]{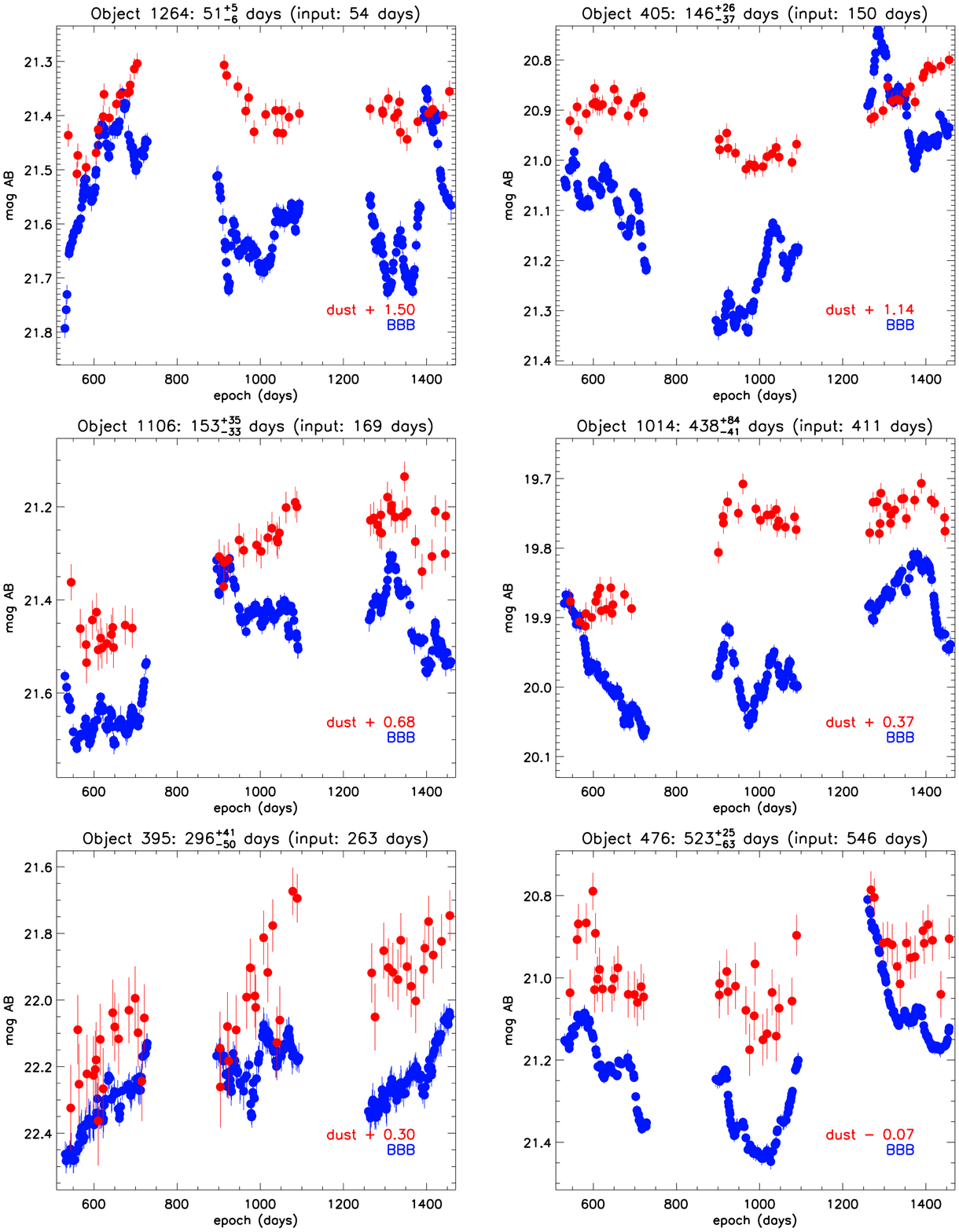}
	\caption{Dust time lag recovery for the six examples in Fig.~\ref{fig:ex_lc}. The blue/darker circles with error bars show the recovered BBB light curve at $0.55\,\micron$ rest-frame wavelength. The red/lighter data points correspond to the recovered hot dust light-curve in the observed $Ks$-band after subtraction of the BBB component, with offsets indicated. As the redshift increases, the contribution of dust emission to the $Ks$-band decreases but lags can still be recovered. The recovered lags (observer frame) with errors and input lags are indicated in the headline of each panel.}
	\label{fig:recov}
\end{figure*}

The standard method to determine dust lags is to calculate the cross-correlation function (CCF) for a range of lags and determine the peak. However, this technique requires well sampled light curves and good S/N ratios; otherwise the CCF will have low values, and it becomes difficult to distinguish between the true lag and spurious peaks. Also, this method is strongly affected by the window function of the observations (seasonal gaps, total survey coverage). Instead, we attempt to recover the time lags with a technique that maximises the use of information within the light curves. For this, we make use of the fact that the dust light curve $F_\mathrm{dust}(t)$ is a rescaled, smeared-out and shifted version of the UV/optical light-curve $F_\mathrm{BBB}(t)$, i.e.
\begin{equation}\label{eq:dtrans}
F_\mathrm{dust}(t) = \int F_\mathrm{BBB}(t^\prime-\tau) \cdot \Phi(t-t^\prime) \ \mathrm{d}t^\prime,
\end{equation}
where $\tau$ is the lag between the hot dust variability and the optical variability, and $\Phi(t)$ is the transfer function, which controls the smearing and rescaling. The key idea behind this approach is that dust reprocesses the UV-peaked BBB emission into infrared photons. This description resembles previous studies of broad emission line lags \citep[e.g.][]{Zu11,Che12,Che13} with the extension that we do parameterise $\Phi(t)$ instead of assuming an \textit{ad hoc} shape (see below). Moreover, as discussed in \citet{Hon14b}, we do not use the full dust and BBB light curves for lag recovery, but only the relative variability $(F(t) - \left<F(t)\right>)/\left<F(t)\right>$ about the mean $\left<F(t)\right>$. This procedure better isolates the variability pattern and the effect of $\Phi(t)$.

A caveat in recovering the dust lags arises from the fact that the observed light curves are not continuous but contain gaps of varying lengths. In order to overcome this problem, we invoke the popular flux randomisation/random subset selection (FR/RSS) method \citep{Pet98}. It uses the variability characteristics of the observed light curves (i.e. typical flux changes for given intervals) to interpolate regularly-spaced random realisations of the light curves based on a Monte Carlo method. In principle, we could use the CAR process invoked to generate the mock light curves to interpolate the mock-observed light curves. This would involve fitting the observed power spectra with the CAR prescription and picking fluxes between observed epochs according to the fitted CAR parameters. However, in this case we would implicitly use our knowledge about the process that generated the variability to recover the light curves, which carries the risk that the errors on the final recovered lags would be underestimated. In addition, recent Kepler observations challenge the CAR process as the driver for accretion variability in AGN \citep{Mus11}, which is why a more data-driven approach to light curve interpolation (such as FR/RSS) will introduce less model dependence for the real survey.

The scheme we follow to model the light curves and recover dust lags starts with determining $F_\mathrm{BBB}(t)$. For that, we make use of the multi-band coverage of the BBB emission in the DES $griz$ filters and, for higher redshift sources, the VISTA $J$ filter. Indeed, we cover the $0.3-0.9\,\micron$ rest-frame region with at least 4 filters in all of the targeted redshift range of $0 < z < 1.2$. In using information from at least four broad filters, we are also less sensitive to potential contamination by emission lines and their variability. In fact, given an object's redshift, it is possible to sub-select an optimal set of filters to minimise broad-line contamination. We resample observations in each band 20 times using the FR/RSS method, with the number of samples limited by computational resources. For each epoch observed by at least one of the $grizJ$ bands, we determine the BBB continuum spectral slope via a power-law fit, determine the flux at a rest-frame wavelength of 550\,nm, and assign errors according to the variation among the randomly resampled light curves. The result is a very high quality light curve of the driver emission for the dust variability (see Fig.~\ref{fig:recov}, blue data points).

In the next step, we recover the dust variability. For this, we start with the FR/RSS-resampled light curves in each of the $grizJ$ bands. We then fit a power law to the resampled $grizJ$ flux at each $Ks$ band epoch and extrapolate the power law to the $Ks$ band. This provides us with the BBB contribution to the $Ks$ band. We then subtract the BBB contribution from the observed $Ks$ data and obtain a clean hot dust light curve, which we will just refer to as the ``dust light curve'' in the following. Recovered BBB and dust light curves for the six example objects in Fig.~\ref{fig:ex_lc} are shown in Fig.~\ref{fig:recov}.

\begin{table*}
\begin{center}
\caption{Comparison of constraints on $\Omega_\Lambda$ from different surveys}\label{tab:lcdm}
\begin{tabular}{l c c c c c l}
\hline
survey & \multicolumn{3}{c}{number of objects} & \multicolumn{2}{c}{$\Omega_\Lambda$ in $\Lambda$CDM} & reference \\
& all & $z>0.3$ & $z>0.7$ & stat. & stat. + syst. & \\ \hline
Union 2.1 type Ia supernovae & 580 & 285 & 90 & $0.723^{+0.022}_{-0.021}$ & $0.705^{+0.043}_{-0.040}$ & \citet{Suz12}\\
JLA type Ia supernovae & 740 & 270 & 108 & $0.711\pm0.018$ & $0.705\pm0.034$ & \citet{Bet14} \\
\veils\ AGN simulations & 437 & 408 & 143 & \multicolumn{2}{c}{$0.718^{+0.043}_{-0.047}$} & this work; input $\Omega_\Lambda = 0.685$ \\ \hline
\end{tabular}
\end{center}
\end{table*}

Next, we take the recovered BBB light curve at 550\,nm rest frame as input to fit eq.~(\ref{eq:dtrans}), after subtracting and normalising by the mean flux, and parametrise the transfer function $\Phi(t)$ in terms of a power-law flux decay in time. For broad lines, the adoption of a top-hat transfer function has become very common. However, there is strong evidence from IR interferometry as well as radiative transfer theory that, at least for the dust, a power law shape is more appropriate \citep[e.g. see][]{Hon11,Hon14b}. Therefore, we fit for a parametrisation of the transfer function 
\begin{equation}
\Phi(t) = \Phi_0 \cdot (t/\tau + 1)^\alpha, 
\end{equation}
where $\tau$ is the time lag, $\alpha$ is the power law index of the brightness distribution, and $\Phi_0$ is the normalisation amplitude to obtain $\int_0^\infty \Phi(t) \mathrm {d}t = 1$.

The BBB light curve is FR/RSS-resampled and a time lag with uncertainty is inferred by fitting the resampled, shifted and normalised BBB light curve to the observed dust light curve via the power-law index of $\Phi(t)$, an offset of the mean magnitude, and a stretch factor, using the Levenberg-Marquardt algorithm implemented in \textit{mpfit} \citep{Mar09}. For each resampled version, the $\chi^2$ space is very uneven with several local minima. To distinguish the various minima objectively and without expectation bias, we redo the fitting for each of the resampled light curves 50 times by selecting random-uniform starting parameters. The most frequent $\tau$ (a "dominant $\chi^2$ minimum") is considered the best-fit value for the particular BBB resample.

\begin{figure*}
	\includegraphics[width=0.9\textwidth]{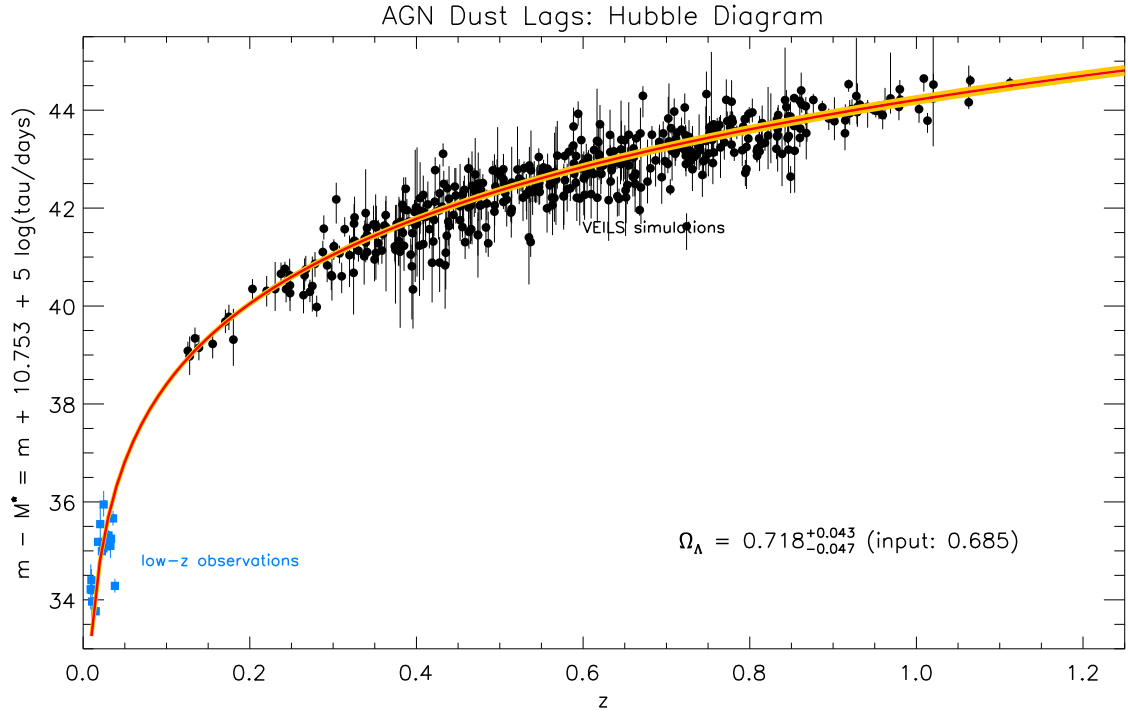}
	\caption{Hubble diagram based on the 437 AGN dust lags recovered in our \veils\ simulations based on 3-year light curves with about 15 observed epochs in the $Ks$ band per 6-month observing season. The red line indicates the best-fit $\Lambda$CDM cosmology with flat geometry and the orange lines represent the 68\% confidence region. The inferred dark energy density parameter is indicated. We also overplot the current low-$z$ sample from literature (blue squared data points) to illustrate how they may be used to help determining the normalisation of the Hubble function (see text for details).}
	\label{fig:hubdia}\end{figure*}

The fitting procedure is repeated for 40 random FR/RSS Monte-Carlo-resampled versions of the BBB light curve. In this process, we also take into account the uncertainties in the recovered dust fluxes and the limited sampling of the dust light curve. Finally, we calculate a mean $\tau$ and 68\% confidence intervals from the distribution of fitted lag values in all FR/RSS samples. This recovered $\tau$ explicitly includes the effects from sparse and inhomogeneous sampling of the dust and BBB light curves.

\section{Results}\label{sec:res}

\subsection{Dust lag recovery statistics}\label{sec:stat}

We applied our dust lag recovery procedure to the 1350 type 1 AGN. Given the survey characteristics, we have about 15 $Ks$-band mock observations per annual observing season, adding up to a total of $\sim$45 observations for each object over the three year \veils\ campaign. Out of the 1350 type 1 AGN, we recovered lags for 437 of the AGN, or 32\%. This number should be considered a conservative approach since we did not follow through with our lag recovery process if the recovered hot dust emission has S/N $<5$ according to the $Ks$ band limit of 22.5\,mag. We also did not attempt to recover a time lag if the expected lag was $>$730 days. This corresponds to a situation where the first observing season in the BBB light curve drives  dust variability in the last observing season. However, some of the final fitted lags may be longer than 730 days because either the expected lag estimate was too short or the fitting procedure settled for a longer lag.

At magnitudes fainter than 22.5\,mag, it becomes increasingly difficult to define a global minimum in $\chi^2$ consistent among all resampled BBB light curves. Similarly, trying to recover longer lags will result in $<$15 observations to be used to fit the dust light curve. We will review if the number of recovered lags can be increased without compromising the quality of recovered lags once real data are available. For the purpose of this paper, the 437 recovered lags form the basis for constraining cosmological parameters.

\subsection{Constraints on cosmology}

In this section, we demonstrate the suitability of AGN to constrain cosmological parameters. The simulations of AGN  light curves are based on a dark energy density $\Omega_\Lambda^\mathrm{in} = 0.685$ and $w^\mathrm{in} = -1$. These are the parameters we aim at recovering by using the distance moduli based on the recovered dust time lags.

\subsubsection{Dark energy density $\Omega_\Lambda$}

In Fig.~\ref{fig:hubdia}, we present a Hubble diagram for the 437 AGN with recovered dust time lags. The plot shows the distance modulus $m_V-M_V^*$, calculated from eq.~(\ref{eq:sc}) for all objects, versus redshift $z$. In addition, we show a best fit Hubble function to the distance moduli based on a flat $\Lambda$CDM cosmology (i.e. no curvature and $w=-1$; see Sect.~\ref{sec:distmod}). The fit has two free parameters: (1) the dark energy density $\Omega_\Lambda$, and (2) the normalisation $k_n$ of the AGN distance moduli. Both parameters were fitted jointly. When marginalising over $\Omega_\Lambda$, we obtain a normalisation of $-2.5 \log k_n = (10.753 \pm 0.044)$\,mag. It should be noted that sensitivity to cosmological model increases with redshift while the normalisation is sensitive to lower redshift. Since both $\Omega_\Lambda$ and $k_n$ are correlated, it is important that observation cover a range of redshifts to reduce the uncertainties in inferred parameters. In this work, we focus on the \veils-only AGN with redshifts $z>0.12$. Further high-quality low-redshift AGN dust time lags are available in literature and may be used as additional input once \veils\ data are available  \citep[currently at least 20 objects;][]{Sug06,Poz14,Kos14,Poz15}. These are shown in Fig.~\ref{fig:hubdia} for comparison.

From our simulated lag measurements, we are able to constrain the dark energy density $\Omega_\Lambda = 0.718^{+0.043}_{-0.047}$ after marginalising over $k_n$. This implies that we recovered the simulation input dark energy density well within the 1$\sigma$ confidence interval and with a precision of 6\%. We can compare these AGN-based constraints to current type Ia supernova samples. \citet{Suz12} presented the Union 2.1 sample, which contains 580 SNe. More recently, \citet{Bet14} compiled 740 type Ia SNe in the Joint Lightcurve Analysis (JLA). A comparison between both samples and the \veils\ AGN simulation constraints on $\Omega_\Lambda$ are shown in Table~\ref{tab:lcdm}. All three samples have comparable errors despite differences in sample sizes. The reason for this is the redshift distribution in the samples. The majority of AGN have redshift $z>0.3$ while the supernovae are strongly biased towards lower redshifts. This makes AGN more sensitive to the changes in the Hubble function when invoking different cosmological parameters. On the other hand, the scatter of AGN around the mean function is higher, which compensates for the higher redshifts. In summary, the quality of recovered $\Omega_\Lambda$ demonstrates the potential of AGN dust lags as standard candles. It should be noted that the remaining intrinsic scatter in the recovered dust time lags and, in turn, distance moduli is moderate: In order to obtain a reduced $\chi^2_\mathrm{r}=1$, we needed to add $\sigma_\mathrm{int}=0.19$\,mag to each object.

\subsubsection{The equation of state in $w$CDM cosmology}

\begin{figure}
	\includegraphics[width=0.48\textwidth]{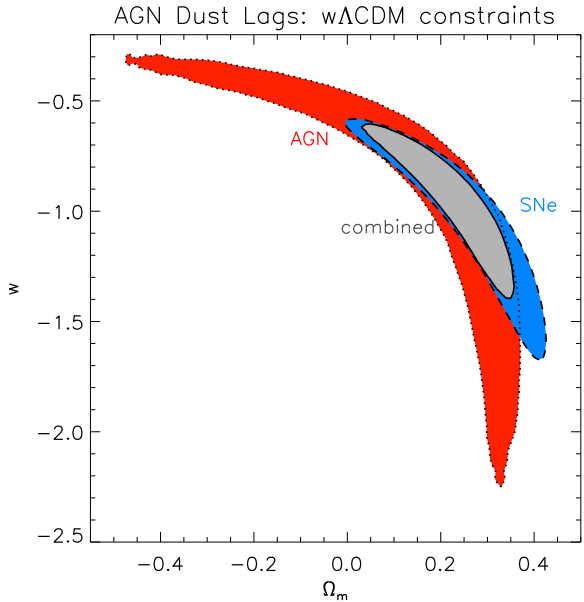}
	\caption{68\% confidence regions for the $w$ and $\Omega_m$ parameters in a flat $w\Lambda$CDM cosmological model. The red area with dotted boundary indicates constraints based on the \veils\ AGN dust time lags only while the blue area with dashed boundary shows the SNe-only constraint based on the Union 2.1 sample. The grey area with solid boundary represents the joint 68\% confidence region.}
	\label{fig:wlcdm}
\end{figure}
 
We further test what new information AGN will contribute to constraining non-standard $w$CDM cosmology. For that, we follow the same fitting procedure as in the previous sub-section, but leave $w$ as a free parameter. In Fig.~\ref{fig:wlcdm}, we show the 68\% confidence interval on combined dark matter density $\Omega_m$ and $w$ constraints for the simulated \veils\ AGN. These are compared to a fit to the Union 2.1 SNe distance moduli from \citet{Suz12}\footnote{We note that the \veils\ AGN simulations have been rescaled for this illustration, so that the input $\Omega_m$ matches the SNe best fit in this parameter.}. The confidence regions are slightly inclined with respect to each other. Since the AGN are at higher redshifts, they are more sensitive to $w$ than current SN data. We also overplot the joint SNe+\veils-AGN 68\% confidence region, which shrinks the combined uncertainty area.

When marginalising over $\Omega_m$, the quality in constraints on $w$ are looser for the simulated \veils\ AGN dust lags ($w = -1.00^{+0.52}_{-0.93}$) than for the Union 2.1 SNe (fit $w = -1.02^{+0.33}_{-0.40}$, including systematics). Improvements are seen when jointly analysing both local standard candles, which results in $w = -0.92^{+0.30}_{-0.36}$. This 10\% reduction in errors is primarily due to the redshift differences between the two object classes, even if the constraints from AGN are weaker than from SNe. However, the major step forward in constraining $w$ involves the joint analysis of these confidence regions with the cosmic microwave background (CMB) and baryonic acoustic oscillations (BAO). The confidence regions of these cosmological probes are essentially perpendicular to the SNe and AGN dust lags.

\section{Discussion of practical challenges in carrying out an AGN dust lag survey}\label{sec:chal}

\subsection{AGN identification, selection, and redshifts}\label{sec:ident}

Using AGN dust lags in cosmology does not require a uniform or well-selected AGN sample. In fact, any AGN that shows sufficient optical and near-IR variability to determine a time lag of the hot dust will suffice. Suitable AGN will be unobscured type 1 AGN with luminosities corresponding to time lags of order a few tens to several hundreds of days, given the flux limit of the survey and the redshift range limit implied by the $Ks$-band filter as the reddest observed waveband. The remaining challenge for such a survey in terms of defining a sample lies in identification of the AGN and obtaining their redshifts. \veils\ will target three well-studied regions with a wealth of multi-band information and catalogues. Given our magnitude limits, we expect that a large fraction/majority of the AGN have been pre-identified. Indeed, the OzDES spectroscopic survey \citep{Yua15} already includes 1700 AGN in the 30 deg$^2$ DES deep fields, which partly overlap with our fields. Further spectroscopic follow-up may be necessary to classify previously unidentified AGN. 

\subsection{Host galaxy subtraction}

The AGN-dust lag method presented here involves 3 observables: The time lag, the redshift, and the apparent magnitude of the AGN. We discussed the lag-recovery strategy (see Sect.~\ref{sec:lag_recov}) and spectroscopic redshifts (see previous section), but it is still necessary to recover the mean AGN magnitude from the underlying host galaxy. While this may be challenging for each individual epoch, our strategy is to combine images from all observed epochs to be able to use well-established host modelling and subtraction tools to disentangle AGN and galaxy on high-S/N multi-band data \citep[e.g.][]{Pen10}. Indeed, given the simultaneous analysis/decomposition of spectroscopic data and DES+\veils\ $grizJKs$ six-band imaging data, we expect to be able to constrain the AGN magnitude to within a fraction of the uncertainty of the lags, which means that we do not expect our cosmological constraining power to be affected by host subtraction. With this multi-dimensional approach, we will also be able to address any potential extinction by adding an extinction component to the decomposition. It is worth mentioning that our simulations build upon the empirical lag-luminosity relation that was obtained under much less ideal conditions than the ones in \veils: The bulk of the currently available data originates from the 2\,m MAGNUM telescope with much lower sensitivity, less angular resolution, and only one optical band \citep[see][and references therein]{Kos14}. We will quantitatively assess the influence of the host galaxies in \veils\ in a future paper once real data are available. However, it is important to note that a systematic bias in any such host-AGN decomposition technique would not affect the constraints on cosmology since this would be absorbed by the normalisation $k_n$ of the whole sample.

A second possibility is using the variable component in each of the $grizJ$ bands as the BBB reference. This circumvents host decomposition and extinction effects, but ignores any non-variable part of the BBB, which may still contribute to dust heating and influence the size of the sublimation radius. If this non-variable fraction is small, it would result in a systematic bias, which will be implicitly compensated for in the normalisation of the Hubble function. On the other hand, if this fraction is luminosity dependent, it would skew the Hubble function. In case of very strong host contamination, the variability of the AGN will be significantly diluted, which increases the measurement errors. We propose to put this to a test with real data, in particular because it has the potential to circumvent any systematic uncertainty from the absolute flux calibration of the images.

\subsection{Evolution of the lag-luminosity relation}

\begin{figure}
	\includegraphics[width=0.48\textwidth]{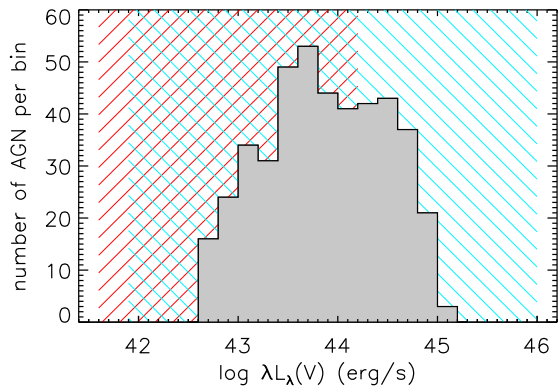}
	\caption{Distribution of $V$-band luminosities of those 437 AGN for which dust time lags were recovered. The red up-hatched area shows the luminosity range covered by low-$z$ dust reverberation mapping while the blue down-hatched area delineates the range covered by near-IR interferometry.}
	\label{fig:hist_lv}
\end{figure}

One unknown in the use of AGN dust time lags as standardisable candles is a potential redshift evolution of the lag-luminosity relation. This evolution may originate from two sources: (1) a change in metallicity, and (2) non-linearity in luminosity. In Sect.~\ref{sec:lag_lum}, we argue that metallicity is probably a minor concern given that we observe the hottest dust with emissivity close to a black body. The major risk of evolution with luminosity comes into play if we reach beyond the luminosity range covered by current low-redshift lag-luminosity studies. Therefore, we compare the range of luminosities for which we recover dust lags in \veils\ to the range established by current low-redshift observations. The the 17 reverberation-mapped objects in \citet{Kos14} span a range of $V$-band luminosities of about $10^{42}\,\mathrm{erg/s} < \lambda L_\lambda(V) < 10^{44}\,\mathrm{erg/s}$. In addition, near-IR interferometry size measurements at low redshift \citep[e.g.][]{Kis09,Kis11a,Kis13} cover approximately $10^{42}\,\mathrm{erg/s} < \lambda L_\lambda(V) < 10^{46}\,\mathrm{erg/s}$.\footnote{When combining $K$-band time lags and interferometry, there is a well known offset in normalisation \citep[e.g][]{Kis07,Kis09,Kos14}. This offset is caused by the underlying dust/brightness distribution and the related sensitivity of the two methods to spatially extended emission \citep[e.g.][]{Hon11}. Modelling the reverberation transfer function allows for compensation of this effect \citep{Hon14b}.} In Fig.~\ref{fig:hist_lv} we compare these ranges with a histogram of $V$-band luminosities for those 437 \veils\ AGN where we recovered dust time lags. The figure shows that \veils\ AGN used for cosmology will cover the same range in luminosity as current low-redshift samples based on which the lag-luminosity relation has been established. This mitigates the risk of luminosity evolution. Nevertheless, we plan to split the observed AGN into sub-samples of different luminosities or metallicities to test if the inferred cosmological constraints differ systematically.

\section{Conclusion and legacy value to future high-redshift AGN cosmology studies}\label{sec:sum}

In this paper, we introduce a new survey that will observationally establish AGN dust time lags as cosmological standardisable candles. We will utilise time-sampled near-IR $J$- and $Ks$-band data of well-studied extragalactic survey fields and combine them with simultaneous multi-band optical data to determine dust time lags of several hundred unobscured AGN. Here, we simulate the survey and evaluate its power in constraining cosmological parameters. We conclude:
\begin{itemize}
\item The new \veils \ public survey with VISTA VIRCAM has the capability to fully establish AGN dust time lags as a new standard candle for cosmology. It is complementary to type Ia supernovae and can assess hidden systematics in the same redshift range targeted by current SNe surveys (e.g. DES).
\item We showed that the constraints on $\Omega_\Lambda$ obtained from the AGN dust lags will be competitive with type Ia supernovae while improvements in $w$ constraints will require a joint AGN+SNe analysis. Indeed, by combining both standard candles, we will be able to improve current $\Omega_\Lambda$ constraints by $\sim$20\% and by $\sim$10\% for $w$. This opens prospects to further narrow down any potential differences between low-redshift and higher redshift cosmological probes (e.g. BAO, CMB).
\item Future efforts in cosmology will focus on higher redshift probes to constrain $w_a$ and $w_0$. The time lag between optical continuum emission in AGN and their broad emission lines will be a key standard candle to reach higher redshifts of $z\sim4$ \citep[e.g.][]{Wat11,Cze13,Kin14}. Since the AGN dust time lags probe exactly the same object class (=unobscured AGN), the dust time lags can provide the low-redshift normalisation for the high-redshift studies. Many of the \veils\ AGN will be monitored with OzDES spectroscopically, which allows for cross-calibration between dust lags and BLR lags and provide the springboard to high redshifts.
\end{itemize}

One of the most substantial advancements will be an immediate growth in the number of low-redshift standard candles when combining SNe with AGN. For that we will need a cross-calibration reference. \citet{Li11} find a SN Ia rate of $0.54\pm0.12$ per century for a Milky Way-type galaxy at $z=0$. Assuming that the Milky Way mass, luminosity, and Hubble type is typical of a Seyfert AGN-hosting galaxy, and considering an increase of the SNe Ia rate by at least a factor of 3 to redshift 0.5 \citep[see also][]{Li11}, we expect that about $5-15$ of the AGN hosts with an established dust lag in \veils\ will display a SN Ia during the life time of the survey. These galaxies will serve as the basis to merge the low-redshift standard candles.

\section*{Acknowledgements}
We are grateful for the comment by the anonymous referee, which helped to improve the manuscript. SFH acknowledges support for this work from the UK Science \& Technology Funding Council (STFC) under grant ST/N000870/1. MK acknowledges support from JSPS under grant number 16H05731. PG acknowledges support from STFC under grant grant ST/J003697/2. KH acknowledges support from STFC grant ST/M001296/1.



%\bibliographystyle{mnras}


%%%%%%%%%%%%%%%%%%%%%%%%%%%%%%%%%%%%%%%%%%%%%%%%%%
\clearpage
\newpage

\appendix

\section{Comparison of simulated lag-luminosity relation to observations}\label{sec:app_taulum}

One goal of this paper is to provide simulations that reflect the observed dust lag-luminosity relation as close as possible to get a realistic estimate of how well cosmological parameters can be constrained. As a sanity check, we test if the scatter in the simulated $\tau-L$ relation is comparable to the real observed one, after processing the mock light curves through our lag recovery pipeline. In the following, $L$ will be expressed as absolute magnitude $M_V$, which is ultimately used in the distance modulus.

\begin{figure}
	\includegraphics[width=0.48\textwidth]{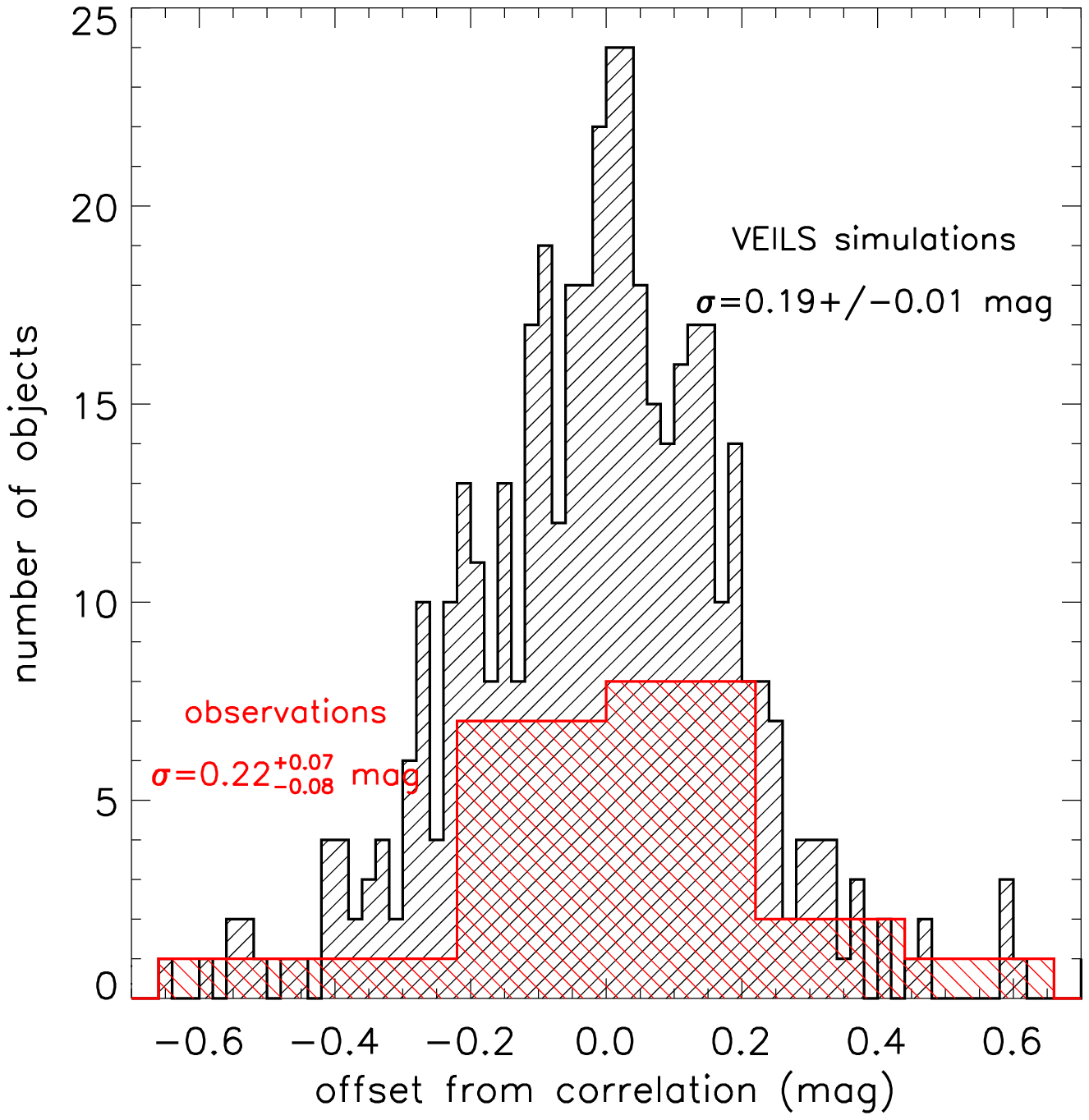}
	\caption{Distribution of inferred time lags from the simulations (black upward-hatched) and observed AGN (red downward-hatched), respectively, around the best-fit lag-luminosity relation. The offsets are expressed in magnitudes as $2.5\log\tau - 2.5\log\tau_\mathrm{pred}$. 68\% confidence intervals are noted for both observed and simulated samples, with the uncertainties inferred from bootstrapping and accounting for the limited number of objects in each sample.}
	\label{fig:hist_taumv}
\end{figure}

We collect $K$-band lags and absolute $V$-band magnitudes $M_V^\mathrm{obs}$ of 20 AGN from the literature \citep{Lir11,Kos14,Poz14,Poz15}. These are used as a comparison sample to our simulations. We then fit the inferred time lags from the mock observations to the absolute magnitudes in the mock observations $M_V^\mathrm{sim} = m_V^\mathrm{sim} - 5\log D_L/\mathrm{10\,pc}$, where $m_V^\mathrm{sim}$ is the mean apparent $V$-band magnitude of the AGN inferred from the multi-band mock light curves. The luminosity distance $D_L$ is calculated from the input redshift given input cosmological parameters. In order to make a comparison to observations, we treat $M_V$ as the independent variable, predict the time lags $\tau_\mathrm{pred}$ by fitting the $M_V-\tau$ relation, and plot the resulting offset of each data point from the best-fit relation in histograms, for observations and simulations respectively. These histograms are shown in Fig.~\ref{fig:hist_taumv} and the widths containing 68\% of the objects are noted. The error on these standard widths have been inferred from bootstrapping the samples to account for the finite number of observed sources in both observations and simulations.

The offsets are shown in magnitudes as $2.5\log\tau - 2.5\log\tau_\mathrm{pred}$ to allow for a direct comparison with the observational uncertainties as well as the distribution in distance moduli. The observed and simulation-recovered lag-luminosity relations are consistent within error bars, with a nominally slightly narrower relations for the simulation-recovered lags. Indeed, it is entirely possible that the slightly narrower distribution in the simulations is driven by the better constraint on the AGN apparent magnitude given that at least 4 bands have been used to infer $m_V$ instead of typically just one band in literature \citep[with the exception of][]{Lir11}. The consistent, and arguably superior, method to determine the time lags, which includes the information from the observed transfer function, may have also improved the scatter in the lags recovered from the simulations. In summary, we conclude that the simulations are properly reflecting the state-of-the-art in dust time lag observations.

%%%%%%%%%%%%%%%%%%%%%%%%%%%%%%%%%%%%%%%%%%%%%%%%%%

% Don't change these lines
\bsp	% typesetting comment
\label{lastpage}
\end{document}